\documentstyle[preprint,eqsecnum,aps]{revtex}

\begin{document}

\draft

\title{Axion-dilaton-modulus gravity theory of Brans-Dicke-type and conformal symmetry}

\author{ Israel Quiros\thanks{israel@uclv.etecsa.cu}}
\address{ Departamento de Fisica. Universidad Central de Las Villas. Santa Clara. CP: 54830 Villa Clara. Cuba }

\date{\today}

\maketitle

\begin{abstract}

Conformal symmetry is investigated within the context of axion-dilaton-modulus theory of gravity of Brans-Dicke-type. A distinction is made between general conformal symmetry and invariance under transformations of the physical units. The conformal degree of symmetry of the theory is studied when quantum fermion (lepton) modes with electromagnetic interaction are considered. Based on the requirement of invariance of the physical laws under general transformations of the units of measure, arguments are given that point at a matter action with non-minimal coupling of the dilaton to the matter fields as the most viable description of the world within the context of the model studied. The geometrical implications of the results obtained are discussed.

\end{abstract}


\section{Introduction}

General relativity (GR) is expected to be the low-energy limit of some underlying quantum theory of gravity\cite{barrow}. String theory seems to represent this underlying theory\cite{gsw}. There are five consistent superstring theories [type I, type IIA, IIB, $E_8\times E_8$ heterotic and SO(32) heterotic] that seem to be related by duality symmetries. The equivalence of these different theories under duality transformations hints at the existence of a more fundamental M-theory\cite{duff}.

Ten-dimensional supergravity theories are the low-energy limit of string theories. In the Neveu-Schwarz/Neveu-Schwarz (NS-NS) bosonic sector these are derivable from the effective action\cite{clw}

\begin{equation}
S=\int d^{10}x\sqrt{-g}\; e^{-\phi}(R+(\nabla\phi)^2-\frac{1}{2}H^2),
\end{equation}
where $R$ is the Ricci scalar of the 10th dimensional spacetime, $\phi$ is the dilaton and $H_{abc}=\partial_{[a}B_{bc]}$ is an antisymmetric tensor field. After compactification to four dimensions Eq. (1.1) can be written as\cite{cew}

\begin{equation}
S=\int d^4 x\sqrt{-g}\; e^{-\psi}(R+(\nabla\psi)^2-n(\nabla\beta)^2-\frac{1}{2}e^{2\psi}(\nabla\sigma)^2),
\end{equation}
where now $R$ is the Ricci scalar of the four dimensional external spacetime, $\psi\equiv \phi-n\beta$ is the shifted dilaton, $n$ is the number of compactified dimensions ($n=6$ in this case), $\beta$ is the modulus field allowing for the variation of the $n$ compactified dimensions, and $\sigma$ is an "axion" field. It can be shown that dimensional reduction from $D+n$ to $D$ dimensions results in a globally $O(n,n;\Re)$ invariant theory in $D$ dimensions\cite{maharana} ($D=4$ in our case). Besides this symmetry action (1.2) is invariant under the transformations of the $SL(2,\Re)$ group\cite{barrow,clw}.

A generalization of the low-energy effective action (1.2) can be given in the form of an effective Brans-Dicke(BD)-type theory of gravity\cite{clw}

\begin{equation}
S=\int d^4 x\sqrt{-g}\; e^{-\psi}(R-(\alpha-\frac{3}{2})(\nabla\psi)^2-\frac{1}{2}e^{\gamma\psi}(\nabla\sigma)^2-\frac{1}{2}(\nabla\beta)^2),
\end{equation}
where the constant $\alpha$ determines the dilaton-graviton coupling\footnotemark\footnotetext{In the bibliography the BD coupling constant $\omega\equiv\alpha-\frac{3}{2}$ is used instead of $\alpha$. The case $\alpha=\frac{1}{2}$ recovers the usual [fundamental string] situation}, while the axion-dilaton coupling is determined by the constant $\gamma$. This action preserves the $O(n,n;\Re)$ and $SL(2,\Re)$ symmetries of (1.2).

Since conformal invariance is a property of string theory at high energies it will be of interest to study the transformation properties of the low-energy action (1.3) in respect to the conformal transformations of the spacetime metric

\begin{equation}
\hat g_{ab}\rightarrow\Omega^2(x) g_{ab},
\end{equation}
where $\Omega(x)$ is a non-vanishing smooth function. The spacetime coincidences [coordinates] do not change under (1.4). This means that the experimental observations [i. e., the verification of the spacetime coincidences] are insensible to the above transformations of the metric tensor. It is a wellknown fact that the transformations (1.4) can be interpreted as point-dependent transformations of the units of length, time and mass\cite{bdk}. 

Although astrophysical observations give evidence against the assumption of a general conformal symmetry of the physical laws\cite{canuto}, in the classical papers \cite{bdk} the authors put forth the requirement that the laws of physics must be invariant not only under general coordinate transformations but, also, under point-dependent transformations of the units of length, time and mass. These transformations are meaningless from the physical point of view. A similar viewpoint was worked out in Ref.\cite{dirac}. In Ref.\cite{iq} this last requirement was raised to a cathegory of a postulate, called therein as Brans-Dicke postulate: {\it The laws of physics, including the laws of gravitation, must be invariant under the transformations of the group of point-dependent transformations of the units of length, time and mass}. Simultaneous holding of the Brans-Dicke postulate and the lack of general conformal symmetry of the physical laws seems to be contradictory. For this reason distinction should be made between general conformal symmetry and invariance under transformations of units. It is precisely the motivation of the present paper.

The paper has been organised in the following way. In Sec. II we discuss the meaning of a conformal transformation of the kind (1.4) in the particular case of axion-dilaton-modulus-vacuum gravity theory of BD-type given by the effective action (1.3). However, any discussion on this subject would be incomplete if we would not take into account the matter fields. Therefore, in Sec. III, we shall study the invariance properties of the action for matter in the form of fermion (lepton) fields with electromagnetic interaction in respect to (1.4). In particular we shall treat the action for quantum electrodynamics (QED) that, in the presence of gravity, can be written as

\begin{equation}
S_{QED}=\int d^4 x \sqrt{-g}(i\;\bar\Psi\gamma^n D_n \Psi-i\;m \bar\Psi\Psi-\frac{1}{4}g^{ns}g^{mr}F_{nm}F_{sr}),
\end{equation}
where $\Psi$ is the Dirac spinor for the fermion (lepton) field, $\gamma^a$ are the Dirac matrixes, $m$ is the mass of the fermion (lepton), and $F_{ab}\equiv A_{a;b}-A_{b;a}$ are the components of the tensor of the electromagnetic field. The covariant derivative $D_a$ is defined as

\begin{equation}
D_a \equiv \partial_a-i\;e A_a,
\end{equation}
where $e$ is the electric charge of the fermion (lepton) field. Since conformal symmetry is not expected to be broken by the presence of quantum matter while it is broken by the presence of classical matter fields\cite{far}, in Sec. IV we study the invariance properties of the general action for classical matter fields. Finally in Sec. V we discuss the geometrical implications of the results obtained in the paper.

\section{Axion-dilaton-modulus-vacuum gravity}

We shall study the transformation properties of the effective action (1.3) in respect to the transformation (1.4). For this purpose we shall write (1.3) as the sum of partial actions

\begin{equation}
S=S_{g\psi}+S_\sigma+S_\beta,
\end{equation}
where $S_{g\psi}$ is the dilaton-graviton part of (1.3)

\begin{equation}
S_{g\psi}=\int d^4 x\sqrt{-g}\; e^{-\psi}(R-(\alpha-\frac{3}{2})(\nabla\psi)^2),
\end{equation}
$S_\sigma$ is the axion part

\begin{equation}
S_\sigma=-\frac{1}{2}\int d^4 x\sqrt{-g}\; e^{(\gamma-1)\psi}(\nabla\sigma)^2,
\end{equation}
and $S_\beta$ is the modulus part of (1.3)

\begin{equation}
S_\beta=-\frac{1}{2}\int d^4 x\sqrt{-g}\; e^{-\psi}(\nabla\beta)^2.
\end{equation}

Under the conformal transformation of the metric (1.4), the dilaton-graviton part of the action (1.3) [Eq. (2.2)] is mapped onto

\begin{equation}
S_{g\psi}=\int d^4 x\sqrt{-g}\; e^{-\psi}\Omega^2(R-6\frac{\Box\Omega}{\Omega}-(\alpha-\frac{3}{2})(\nabla\psi)^2),
\end{equation}
meanwhile (2.3) is mapped onto

\begin{equation}
S_\sigma=-\frac{1}{2}\int d^4 x\sqrt{-g}\;\Omega^2 e^{(\gamma-1)\psi}(\nabla\sigma)^2,
\end{equation}
and (2.4) is mapped onto

\begin{equation}
S_\beta=-\frac{1}{2}\int d^4 x\sqrt{-g}\;\Omega^2 e^{-\psi}(\nabla\beta)^2.
\end{equation}

We set now

\begin{equation}
\Omega^2 = e^{\tau\psi},
\end{equation}
where $\tau$ is a constant parameter. The cases $\tau=1$ and $\tau\neq 1$ we shall study separatelly since, as we shall see below, they represent very different situations.

In general, $\frac{\Box\Omega}{\Omega}=\frac{1}{2}\tau\Box\psi+\frac{1}{4}\tau^2(\nabla\psi)^2$ so when $\tau=1$ the partial actions (2.5), (2.6) and (2.7) can be written as

\begin{equation}
S_{g\psi}=\int d^4 x\sqrt{-g}(R-\alpha(\nabla\psi)^2),
\end{equation}

\begin{equation}
S_\sigma=-\frac{1}{2}\int d^4 x\sqrt{-g}\;e^{\gamma\psi}(\nabla\sigma)^2,
\end{equation}
and

\begin{equation}
S_\beta=-\frac{1}{2}\int d^4 x\sqrt{-g}\;(\nabla\beta)^2,
\end{equation}
respectively. In other words, transformation (1.4), (2.8) with $\tau=1$ maps the string-frame action (1.3) into the Einstein frame where the dilaton is minimally coupled to curvature. In this case the effective action (1.3) is not invariant under the conformal transformation (1.4). Hence, since the physical laws must be invariant under point-dependent transformations of the physical units\cite{bdk}, we may conclude that transformation (1.4), (2.8) with $\tau=1$ can not be interpreted properly as a transformation of the units of length, time and mass\cite{iq}.

In the case when $\tau\neq 1$ the dilaton-graviton action (2.2) can be written as

\begin{equation}
S_{g\psi}=\int d^4 x\sqrt{-g}\; e^{(\tau-1)\psi}(R-(\alpha-\frac{3}{2}(\tau-1)^2)(\nabla\psi)^2),
\end{equation}
so if we redefine the dilaton

\begin{equation}
\psi\rightarrow\frac{\psi}{1-\tau},
\end{equation}
and, at the same time, allow for a parameter transformation

\begin{equation}
\alpha\rightarrow\alpha(1-\tau)^2,
\end{equation}
hence, finally

\begin{equation}
S_{g\psi}=\int d^4 x\sqrt{-g}\; e^{-\psi}(R-(\alpha-\frac{3}{2})(\nabla\psi)^2).
\end{equation}

Therefore, the dilaton-graviton action (2.2) is invariant under (1.4), (2.8), (2.13) and (2.14) with $\tau\neq 1$. It is a very nice feature of this part of the effective action (1.3). Under these transformations the partial actions (2.3) [see Eq. (2.6)] and (2.4) [see Eq. (2.7)] can be written as

\begin{equation}
S_\sigma=-\frac{1}{2}\int d^4 x\sqrt{-g}\; e^{(\frac{\gamma}{1-\tau}-1)\psi}(\nabla\sigma)^2,
\end{equation}
and

\begin{equation}
S_\beta=-\frac{1}{2}\int d^4 x\sqrt{-g}\; e^{-\psi}(\nabla\beta)^2.
\end{equation}
respectively. Therefore, the modulus part of the effective action (1.3) is already invariant under (1.4), (2.8), (2.13) and (2.14) with $\tau\neq 1$. If we set, also, the parameter transformation

\begin{equation}
\gamma\rightarrow\gamma(1-\tau),
\end{equation}
hence the axion part of the action (1.3) [Eq. (2.3)] is invariant too.

The set of transformations (1.4), (2.8), (2.13), (2.14) and (2.18) with $\tau\neq 1$ is a one-parameter group of transformations. In fact, a composition of two successive transformations with parameters $\tau_1 \neq 1$ and $\tau_2 \neq 1$ yields a transformation of the same kind with parameter $\tau_3=\tau_1+\tau_2-\tau_1\tau_2 \neq 1$, such that $\tau_3(\tau_1 ,\tau_2)=\tau_3 (\tau_2 ,\tau_1)$ and the group is commutative. The identity of this group is the transformation with $\tau=0$, while the inverse of a transformation with parameter $\tau$ is a transformation with parameter $\bar\tau=\frac{\tau}{\tau-1}$. A particular case of interest is when $\tau=2$, i. e., $g_{ab}\rightarrow e^{2\psi} g_{ab}$, $\psi\rightarrow -\psi$, $\alpha\rightarrow\alpha$ and $\gamma\rightarrow -\gamma$. In this case, since the string coupling ${\it g} \sim e^\psi$, the conformal transformation with $\tau=2$ interchanges strong and weak coupling regimes.

In Ref.\cite{iq} arguments were given that point at transformations (1.4), (2.8) and (2.13) with $\tau\neq 1$ as a serious candidate for the so claimed group of point-dependent transformations of the units of length, time and mass. In this sense, it is very encouraging that the effective action (1.3) is invariant under the transformations (1.4), (2.8), (2.13), (2.14) and (2.18) with $\tau\neq 1$. In what follows we shall call this group of transformations as "group of units transformations". Therefore, since any consistently formulated physical law must be invariant under the transformations of this group, henceforth we shall ask for invariance of the laws of physics (including the laws of gravity) under the transformations (1.4), (2.8), (2.13), (2.14) and (2.18) with $\tau\neq 1$.

In view of the partial results of this section, it should be remarked that, although the laws of gravity that are derivable from the action (1.3), in general, are not invariant under (1.4), (2.8) and (2.13) when $\tau$ is arbitrary [this includes the case when $\tau=1$], they are invariant under these transformations [plus (2.14) and (2.18)] when $\tau\neq 1$, i. e., they are invariant under the transformations of the "group of units transformations".

\section{The action for quantum electrodynamics}

The effective action (1.3) considers only the vacuum case, i. e., no ordinary matter fields besides the dilaton, axion and modulus fields are present. Therefore, it will be of interest to look at more general actions that include some kind of ordinary matter. In this sense we first shall study matter in the form of an electromagnetic [radiation] field given by

\begin{equation}
S_{em}=-\frac{1}{4}\int d^4 x \sqrt{-g}g^{ns}g^{mr}F_{nm}F_{sr},
\end{equation}
where $F_{ab}\equiv A_{a;b}-A_{b;a}=A_{a,b}-A_{b,a}$ is the tensor of the electromagnetic field. Suppose that under (1.4)

\begin{equation}
A_c\rightarrow A_c-\frac{i}{e}\Omega^{-1}\Omega_{,c}.
\end{equation}

It can be easily checked that under (1.4) [arbitrary functional form of $\Omega^2(x)$] and (3.2) $F_{ab}\rightarrow F_{ab}$, i. e., the electromagnetic tensor is unchanged by these transfromations. In fact, under (1.4), $P^a = m\frac{dx^a}{ds}\rightarrow\Omega^{-2}P^a$ so $P_a = g_{an}P^n\rightarrow P_a$ [it is unchanged by (1.4)]. This means that the derivative is unchanged too, $\partial_a\rightarrow\partial_a$ [recall consistency of the quantum relation $P_a\rightarrow i\partial_a$]. Therefore, the action (3.1) of the electromagnetic field is invariant under (1.4) and (3.2). The particular functional form of the conformal factor $\Omega^2(x)$ is not important in this case. This result is, of course, not a new one. It is wellknown that the presence of ordinary matter with a traceless stress-energy tensor [radiation] does not break the conformal invariance of the action theory (1.3)\cite{far}. 

The next step is to include quantum matter with electromagnetic interaction. It is expected that quantum matter does not break the conformal invariance of the theory too. We shall study the action for quantum electrodynamics (QED) Eq. (1.5) when the transfromation (1.4) is performed. Under this transformation the Dirac matrixes transform like

\begin{equation}
\gamma^a\rightarrow\Omega^{-1}\gamma^a,
\end{equation}
while the spinor of the fermion (lepton) states

\begin{eqnarray}
\Psi &\rightarrow &\Omega^{-1}\Psi,\;\;\Psi^+\rightarrow\Omega^{-1}\Psi^+\nonumber\\
\bar\Psi&\equiv &\gamma^0\Psi^+\rightarrow\Omega^{-2}\bar\Psi.
\end{eqnarray}

This means, in particular, that under (1.4) $\rho\sim\bar\Psi\gamma^0\Psi\rightarrow\Omega^{-4}\rho$ so the probability

\begin{equation}
W=\int d^4x\sqrt{-g}\rho,
\end{equation}
is unchanged by (1.4) too. The transformation (3.4) can be interpreted geometrically as a local transformation of the unit of measure of the probability amplitude.

It can be checked without difficulty that the action (1.5) for QED is invariant under (1.4), (3.2-3.4).\footnotemark\footnotetext{We have taken into account that under (1.4) the mass of the fermion (lepton) states transforms as $m\rightarrow\Omega^{-1}m$} In a more general fashion, the action (1.5) is invariant under the "gauge" transformations (1.4) with an arbitrary functional form of $\Omega^2(x)$ [in particular $\Omega^2=e^{\tau\psi}$, $\tau$ arbitrary],

\begin{equation}
\Psi\rightarrow e^{i e \pi (x)}\Psi,
\end{equation}
and

\begin{equation}
A_c\rightarrow A_c -\pi (x)_{,c},
\end{equation}
where $\pi (x)\equiv\mu (x)+\frac{i}{e}\ln \Omega (x)$ [$\mu$ is an arbitrary realvalued function] is a complex function. Therefore, the action (1.5) is invariant under the local non-unitary group $nU(1)$ with element $e^{i e \pi (x)}$. This means that under (3.6) the norm of the fermion (lepton) states is not preserved locally. However, it is not catastrophic since (3.6) is accompained by a change in the geometry. In effect, under (1.4), what appears as a manifold of Riemann structure is mapped onto a manifold of conformally-Riemannian [Weyl-integrable] structure\cite{iq,novello} in which the units of measure (including now the unit of measure of the probability amplitude) is point-dependent. This point will be discussed in detail in Sec. IV and in Sec. V.

Summing up. The effective action

\begin{eqnarray}
S_{eff}=\int d^4 x\sqrt{-g}\; &\{&e^{-\psi}(R-(\alpha-\frac{3}{2})(\nabla\psi)^2-\frac{1}{2}e^{\gamma\psi}(\nabla\sigma)^2-\frac{1}{2}(\nabla\beta)^2)+\nonumber \\ &i&\;\bar\Psi\gamma^n D_n \Psi-i\;m \bar\Psi\Psi-\frac{1}{4}g^{ns}g^{mr}F_{nm}F_{sr}\},
\end{eqnarray}
is invariant under the transformations (1.4), (2.8), (2.13), (2.14), (2.18) with $\tau\neq 1$, together with the transformations (3.2-3.4), i. e., it is invariant under the [extended] group of units transformations [it includes now transformation of the probability amplitude, electric and magnetic units of measure]. 

We recall that, although the laws of QED are preserved by (1.4) with $\Omega^2$ arbitrary, including $\Omega^2=e^\psi$, this last transformation may not be interpreted properly as a units transformation since it changes the laws of gravity that are derivable from (1.3). Transformation (1.4) with $\Omega^2=e^\psi$ is just a conformal transformation that allows "jumping" from one formulation of the theory into its conformal. Both conformal formulations are observationally equivalent as noted before. QED is therefore conformally invariant in a general sense ($\Omega^2(x)$ is an arbitrary function).

\section{Matter fields and conformal symmetry}

The next step is to introduce classical matter fields. The matter part of the action can then be written as

\begin{equation}
S_{matter}=16\pi\int d^4x\sqrt{-g}\;L_{matter},
\end{equation}
where there is no coupling between the dilaton and the matter fields [minimal coupling].

The equation of motion for an uncharged and spinless particle that is derivable from Eq. (4.1) 

\begin{equation}
\frac{d^2 x^a}{ds^2}+\{^{\;\;a}_{mn}\}\frac{dx^m}{ds}\frac{dx^n}{ds}=0,
\end{equation}
where $\{^{\;a}_{bc}\}\equiv\frac{1}{2}g^{an}(g_{bn,c}+g_{cn,b}-g_{bc,n})$ are the Christoffel symbols of the metric, coincides with the equation defining time-like geodesic curves on a spacetime of Riemann configuration. For this reason theories with minimal coupling of the matter fields to the metric are naturally linked with Riemann geometry.

Under the conformal transformation (1.4) the action (4.1) is mapped into $S_{matter}=16\pi\int d^4 x\sqrt{-g}\;\Omega^4 L_{matter}$, so it changes under this transformation independent of the specific functional form of the conformal factor $\Omega^2(x)$. The same is true for the Eq. (4.2). This was expected since the presence of classical matter fields is supposed to break the conformal invariance of the theory. However, since the transformation (1.4) with $\Omega^2=e^{\tau\psi}$, $\tau\neq 1$ may be properly interpreted as a point-dependent transformation of the units of length, time and mass and, since the physical laws must be invariant under these transformations, hence we may conclude that minimal coupling of the matter fields is not a viable coupling. Therefore we shall look at a matter action with non-minimal coupling that is invariant under (1.4), (2.8) and (2.13) with $\tau\neq 1$. For this purpose we take an action of the form

\begin{equation}
S_{matter}=16\pi\int d^4x\sqrt{-g}\;e^{a\psi}L_{matter},
\end{equation}
where $a$ is some constant factor to be specified later, and the non-minimal coupling of the dilaton to the matter fields is made evident through the exponent of the dilaton under the integral in Eq. (4.3). Under (1.4), (2.8) and (2.13) with $\tau\neq 1$, the action (4.3) is mapped into

\begin{equation}
S_{matter}=16\pi\int d^4x\sqrt{-g}\;e^{(\frac{2\tau+a}{1-\tau})\psi}L_{matter}.
\end{equation}

Invariance of (4.3) under the above transformation requires fulfillment of the following equality

\begin{equation}
a=\frac{2\tau+a}{1-\tau}.
\end{equation}

Eq. (4.5) is satisfied [for any $\tau\neq 1$] only if $a=-2$, i. e., the matter action we are looking for is

\begin{equation}
S_{matter}=16\pi\int d^4x\sqrt{-g}\;e^{-2\psi}L_{matter}.
\end{equation}

This action is invariant under the transformations of units studied here. The equation of motion that is derivable from Eq. (4.6) for an uncharged and spinless particle\cite{iq}

\begin{equation}
\frac{d^2 x^a}{d\hat s^2}+\{^{\;\;a}_{mn}\}_{hat}\frac{dx^m}{d\hat s}\frac{dx^n}{d\hat s}-\frac{\psi_{,n}}{2}(\frac{dx^n}{d\hat s}\frac{dx^a}{d\hat s}-\hat g^{na})=0,
\end{equation}
is, consequently, invariant too under (1.4), (2.8) and (2.13) with $\tau\neq 1$.

The action (4.6) is conformal to the action (4.1) under (1.4) with $\Omega^2=e^\psi$ [the same is true for the equations (4.7) and (4.2), they are conformal to each other under (1.4) and (2.8) with $\tau\neq 1$]. Therefore, theories with non-minimal coupling of the dilaton to the matter fields of the kind (4.6) are naturally linked with conformally-Riemann geometry often acknowledged as Weyl-integrable geometry (WIG)\cite{iq}. It is a special case of generic Weyl geometry\cite{weyl} which allows point-dependent length of vectors to be integrable along a closed path. In effect, in a Weyl-integrable geometry the covariant derivative of the metric tensor $g_{ab}$ is non-vanishing

\begin{equation}
g_{ab;c}=\psi_{,c}\;g_{ab},
\end{equation}
and the gauge [Weyl] vector coincides with the gradient of the dilaton $\psi_{,c}$.\footnotemark\footnotetext{In Eq. (4.8) covariant derivative is defined in a general affine sense, i. e., through the Weyl affine connection $\Gamma^a_{bc}=\{^{\;a}_{bc}\}-\frac{1}{2}(\psi_{,b}\;\delta^a_c+\psi_{,c}\;\delta^a_b-g_{bc} g^{an}\;\psi_{,n})$} In this case the integral of the change of the length of a given vector $V^a(x)$ ($dl=l\;dx^n\psi_{,n}$ with $l\equiv\hat g_{nm} V^n V^m$) along a closed path vanishes: $\oint dl=0$. This means that the undesirable ocurrence of the "second clock" effect is overcome\cite{novello}. Under the conformal transfromation (1.4) with $\Omega^2=e^\psi$ the requirement (4.8) is transformed into

\begin{equation}
g_{ab;c}=0,
\end{equation}
where now the covariant derivative is given in terms of the Christoffel symbols of the metric. Eq. (4.9) is the basic requirement of Riemann geometry. Therefore, we have shown that WIG is conformal to Riemann geometry under (1.4) with $\Omega^2=e^\psi$.

The main lesson to be learned here is that, following the requirement of invariance under transformations of the physical units, only theories with non-minimal coupling of the dilaton to the matter fields of the kind (4.6) have chance of success. For this reason we propose the effective action

\begin{eqnarray}
S_{eff}=\int d^4 x\sqrt{-g}\; e^{-\psi}&(&R-(\alpha-\frac{3}{2})(\nabla\psi)^2-\frac{1}{2}e^{\gamma\psi}(\nabla\sigma)^2-\nonumber\\ &\frac{1}{2}&(\nabla\beta)^2+16\pi\;e^{-\psi}L_{matter}),
\end{eqnarray}
to represent the low-energy limit of string theory. In Ref. \cite{iq,qbc} the issue of the spacetime singularities has been treated within the context of an action of the kind (4.10) with constant axion and modulus fields. It has been shown there that the spacetime singularities that are inherent to spacetimes of Riemann structure may be avoided if the geometrical interpretation of the laws of gravity is to be given on the grounds of Weyl-integrable geometry.

\section{Astrophysical observations and the geometry of the world} 

In the former section it has been shown that, under the conformal transformation (1.4) with $\Omega^2=e^\psi$ (Eq. (2.8) with $\tau=1$), WIG that is given by the basic requirement Eq. (4.8), is transformed into Riemann geometry that is given by the basic requirement Eq. (4.9) that is conformal to (4.8). Therefore, the nature of the spacetime manifold [its underlying geometry] is not preserved under the above conformal transformation. This was expected since the laws of gravity change under this conformal transformation. In effect, under (1.4) with $\Omega^2=e^\psi$, the action (4.10) is mapped into the Einstein frame where it looks like

\begin{equation}
S=\int d^4 x \sqrt{-g}(R-\alpha(\nabla\psi)^2-\frac{1}{2}e^{\gamma\psi}(\nabla\sigma)^2-\frac{1}{2}(\nabla\beta)^2+16\pi L_{matter}).
\end{equation}

Although, supposedly, astrophysical observations raise questions respecting the assumption of a general conformal symmetry of physical laws\cite{canuto}, the gauge invariance of measuring units suggested by Dicke\cite{bdk} and worked out by Canuto et al.\cite{canet} is hardly questionable based on observational arguments only. In effect, we have shown that, in general, the laws of physics [including the laws of gravity] change under the conformal transfromation $g_{ab}\rightarrow e^\psi g_{ab}$. We have shown this for a dilaton-axion-modulus-graviton model. However, as already remarked at the beginning of this section, together with the change in the formulation of the physical laws, under the above transformation of the metric, the geometrical structure of the manifold where the physical laws are to be interpreted changes too. The point is that the conformal transformation of the metric of the kind discussed here does not affect the spacetime coincidences [coordinates]\cite{bdk}. Therefore, the spacetime measurements -being nothing but just verifications of the spacetime coincidences- are not affected by such a conformal transformation\cite{iq,qbc}. In other words, astrophysical observations are unable to differentiate a spacetime of Riemann structure from a Weyl-integrable spacetime.

Nevertheless, symmetry arguments can be approached in order to resolve this "duality" of the geometrical interpretation of the physical laws\cite{iq,qbc}. In the former sections it has been shown that the low-energy effective action (4.10) is invariant under the group of transformations of the physical units [Eq. (1.4), (2.8), (2.13), (2.14) and (2.18) with $\tau\neq 1$]. Since it is a natural requirement any consistently formulated physical law must share, we propose the action (4.10), in which the underlying manifold is of Weyl-integrable nature, to represent the low-energy limit of string theory -the final theory of spacetime. This means that, in spite of the observational equivalence of Riemannian and Weyl-integrable spacetimes, symmetry arguments hint at a Weyl-integrable geometry as the geometry of the world\cite{iq}.

Put in other words. Since the laws of gravity as formulated in the string-frame (action (4.10)) are invariant under the transformations of the "group of units transformations", Weyl-integrable geometry, to which this formulation of gravitation is naturally coupled (see Sec. IV), is expected to possess the above group of symmetry. It is precisely the case. In fact, under (1.4), (2.8) and (2.13) with $\tau\neq 1$, the requirement (4.8) of WIG is mapped into

\begin{equation}
g_{ab;c}=(1-\tau)\psi_{,c} g_{ab}
\end{equation}
since, under $g_{ab}\rightarrow e^{\tau\psi}g_{ab}$; $g_{ab;c}\rightarrow (\tau\psi_{,c} g_{ab}+g_{ab;c})e^{\tau\psi}$. Therefore, under $\psi\rightarrow\frac{\psi}{1-\tau}$ (Eq. (2.13)), Eq. (5.2) is mapped back into (4.8). In the same way, it can be shown that Eq. (4.7), being the geodesic equation of Weyl-integrable geometry\cite{iq}, is invariant under (1.4), (2.8) and (2.13) with $\tau\neq 1$. These results lead us to conclude that the Weyl-integrable structure of spacetime is preserved under the transformations of the "group of units transformations". Unfortunately this is not true for spacetimes of Riemann structure. Therefore, symmetry considerations lead us to consider that a Weyl-integrable configuration is more viable than a Riemannian one. Other arguments pointing at this direction we can find in Ref. \cite{novello} and references therein.

When quantum electrodynamics is concerned, we have shown that the action (1.5) for QED is invariant in respect to general conformal transformation (1.4) with an arbitrary functional form of $\Omega^2(x)$ [in particular $\Omega^2=e^{\tau\psi}$ with $\tau$ arbitrary]. This means that the laws of QED, when formulated in the presence of gravity, look the same in a Riemannian spacetime and in a manifold of Weil-integrable configuration. Conformal invariance in the general sense is a very nice feature of this theory.

A final remark about the transformations (1.4), (2.8), (2.13), (3.2-3.4) with $\tau\neq 1$. These can be viewed as a transformation in the space of the parameters of the theory ($\alpha,\gamma$). In effect, under the above transformations we move within the parameter space ($\alpha,\gamma$) so the set of spacetimes $\{({\cal M},g^{(\alpha,\gamma)}_{ab},\psi^{(\alpha,\gamma)})/\alpha,\gamma\in\Re\}$, where $\cal M$ is a smooth manifold of Weyl-integrable configuration, forms an equivalence class. The spacetimes in this set are physically equivalent for the representation of the physical reality. This equivalence includes, of course, observational equivalence. The physical laws look the same in all of the spacetimes that belong to the former equivalence class.

We propose, in the future, to look at the consequences the formulation of string theory on a manifold of Weyl-integrable nature leads to. This may represent a new alternative for addressing some basic questions of string cosmology. For instance, the cosmological singularity problem, the problem with the observational constrain $\omega>500$ ($\alpha>500$), etc\cite{iq}.

I acknowledge useful conversations with colleagues Rolando Cardenas and Rolando Bonal.

\end{document}